\documentclass[11pt]{article}
\usepackage[margin=1in]{geometry}
\usepackage{amsmath}
\usepackage{xspace}
\usepackage{cite}
\usepackage{url}

\newcommand{\ie}{\emph{i.e.,}\xspace}
\newcommand{\eg}{\emph{e.g.,}\xspace}

\newcommand{\itembf}[1]{\item \textbf{#1}}
\newcommand{\pnew}{\ensuremath{p_{\mbox{New}}}}
\newcommand{\pf}{\ensuremath{p_{\mbox{fp}}}}

\newcommand{\token}{\texttt{tok}}
\newcommand{\tick}{\texttt{tick}}
\newcommand{\user}{\texttt{U}}

\begin{document}
\date{\today}

\title{Anonymous Collocation Discovery:\\ Harnessing Privacy to Tame the Coronavirus\thanks{Author order is alphabetic.}
}

\author{Ran Canetti\thanks{canetti@bu.edu, supported by NSF Grants
1414119 and 1801564, the IARPA Hector program,  and the DARPA SIEVE program.} \hspace{1cm} Ari Trachtenberg\thanks{trachten@bu.edu,  supported, in part, by NSF Grant 1563753.} \hspace{1cm} Mayank Varia\thanks{varia@bu.edu, supported by NSF Grants 1718135, 1915763, and 1931714 and the DARPA SIEVE program.} \\ \\ Boston University }

\date{\today}

\maketitle
\abstract{
Successful containment of the Coronavirus pandemic rests on the ability to quickly and reliably identify those who have been in close proximity to a contagious individual.   Existing tools for doing so rely on the collection of exact location information of individuals over lengthy time periods, and combining this information with other personal information.  This unprecedented encroachment on individual privacy at national scales has created an outcry and risks rejection of these tools.

We propose an alternative: an extremely simple scheme for providing fine-grained and timely alerts to users who have been in the close vicinity of an infected individual. Crucially, this is done while preserving the anonymity of all individuals, and without collecting or storing  any personal information or location history. Our approach is based on using short-range communication mechanisms, like Bluetooth, that are available in all modern cell phones. It can be deployed with very little infrastructure, and  incurs a relatively low false-positive rate compared to other collocation methods. We also describe a number of extensions and tradeoffs.  

We believe that the privacy guarantees provided by the scheme will encourage quick and broad voluntary adoption.  When combined with sufficient testing capacity and existing best practices from healthcare professionals, we hope that this may significantly reduce the infection rate.
}

\vspace{0.7cm}
\begin{center}
\noindent\fbox{\parbox{5.5in}{
To avoid confusion, we stress that this work does \emph{not} propose any direct medical treatment. Rather, it proposes a way to pool together information from the community in order to help (a) direct medical personnel in how to best allocate and use testing resources,  and (b)  direct individuals as to when to get tested and self-quarantine.  
}}
\end{center}
\newpage

\section{Introduction}
The COVID-19 coronavirus has been spreading throughout much of the world at an exponential rate~\cite{zhang2020estimation},
and, though there are promising potential treatments, there are currently no reliable and lasting
mitigations. This suggests that the virus will continue to spread, until either
(i) the relevant region develops herd immunity ($\approx$ 50\% infection among the population~\cite{MGHrounds20}), or (ii) a viable vaccine is developed (12 months away by optimistic estimates~\cite{korman_2020}).

In the meantime, countries around the world are scrambling to slow the rate of infection, mainly to contain the surge of patients needing emergency care that has been overwhelming medical systems. Indeed, protecting the viability of the medical system is critical, both for keeping down the mortality rate among COVID-19 patients, and more fundamentally for preserving the ability of society to function as a whole.
Two types of mechanisms are used to minimize infection rates: \begin{description}
    \item[General Quarantine:]
    Instructing all people in the community to self isolate and severely restricting their movements.
    \item[Targeted Quarantine:] Timely discovery and isolation of infected and potentially infected individuals.
\end{description}
 
Without an existing infrastructure, general quarantine is the simpler approach to implement on short notice. However it has devastating effects both to the lives of individuals (especially those with economically vulnerable backgrounds)  
and to national economies more generally. Furthermore, its effectiveness relies on complete cooperation from all segments of the population, something that becomes increasingly challenging over time. Indeed, this methods raises ethical issues and its effectiveness is not clear~\cite{G20,sars-quarantine}.

The targeted quarantine approach, on the other hand, shows promise for keeping the pandemic contained while restricting the larger
effects on the  population at large.  However, to be effective this approach requires:
\begin{enumerate}
    \item 
    Extensive testing in order to discover contagious individuals as early as possible.
    \item
    An effective and timely mechanism for informing and isolating individuals that were in close proximity to a contagious individual. This is made more challenging by evidence that people become contagious before they are symptomatic, and that surfaces may remain contagious for hours after contact. ~\cite{nejm2020,newman_2020,COVIDsurfaces20}.
\end{enumerate}
Indeed, in countries that have managed to implement these two elements successfully (such as South Korea, Taiwan or Singapore), infection rates have remained relatively low, with correspondingly milder economic and social disruption. 

However, the countries that implemented the targeted isolation mechanism did so at high cost to the privacy of their citizens.  Their solutions required centrally cross-referencing precise cellphone location histories of \emph{all} their citizens against
the identities of infected individuals.  This potentially deeply personal data has then been released with only \emph{ad hoc} partial ``anonymization'' measures that have left the private information of both infected and non-infected individuals vulnerable. Indeed, these characteristics have raised concerns and objection to deployment in other countries and especially in the United States (see e.g. \cite{reutersmarch20,markey-letter}).

\subsection{This work}
The purpose of this work is to demonstrate that it is possible to promptly and reliably notify individuals of past or present collocation with an infected person or surface --- with only minimal compromise of individual privacy and without maintaining any  database of infected individuals or their locations. 

Specifically, we present a cellphone-based scheme that provides individuals with fine-grained, reliable, and timely information on contagion risk due to collocation with an infected individual or surface. 
This is done without disclosing to anyone the identity of the collocated individual(s). Our scheme has the following features:

\begin{itemize}
\itembf{Open participation.} Participation can vary freely over time with users joining and leaving the system at will.

\itembf{Simplicity.} The basic mechanism is \emph{dead simple} to use and understand. While additional features add some complexity, they do so in gradual manner and without interfering with the simplicity of the basic mechanism.  The simplicity may also translate into a more easily implemented and verified solution.

\itembf{Decentralization.}
The scheme maintains a small database on each local cellphone and a database stored on a central registry. However, each database contains no meaningful personal information in and of itself. Furthermore, the aggregation of several databases contains only limited information on other users.

\itembf{Easy deployment.}
The scheme is relatively easy to deploy and requires only minimal infrastructure.

\end{itemize}

We emphasize that being in close proximity with an infected person or surface is \emph{not} a definitive indicator of transmission of disease: two people may be in close proximity without transmitting the virus, or the virus may transfer through aerosols between two parties whose collocation is separated by time.  As such, the proposed ideas are intended to complement (not substitute!) effective prevention techniques such as social/physical distancing \cite{who-advice-for-public}.

\subsection{The basic approach}

We first note that, though GPS data is the natural tool for localization, it can be a relatively poor mechanism for determining {\em close collocation} of individuals, especially in densely populated environments.
Indeed, GPS effectiveness is broadly influenced by a number of factors, such as building density and atmospheric conditions.
In indoor environments, GPS also has trouble effectively discerning vertical separation (say different floors of a building)
or horizontal separation through walls.  As such, collocation information generated by GPS signal alone is bound to have a high rate of false positives, and it is thus typically combined in practical implementations with additional localization sources, further degrading the privacy of participants.   

In contrast, short-range communication technologies, such as Bluetooth Low Energy are more conducive to precisely determine device collocation. Indeed, for some devices it is possible to reduce transmission power of these technologies in order
to limit commercial reception to ranges as short as 2 meters. 
Another attractive aspect of short-range communication technologies is that they are naturally decentralized:  information is exchanged directly between collocated devices without any central intervention. 

This naturally leads to the following basic approach:  each participating cellphone constantly broadcasts, on its short-range communication devices, a random number (token) that changes every few minutes; simultaneously, each phone records the random tokens received from neighboring phones.  As soon as a cellphone is informed that its owner is infected and was potentially contagious during a certain time period, it  uploads to a public registry the tokens that it  transmitted during that time period.
Other participants can then match the tokens they collect against the public registry. If a match is found, the owner knows to get tested as soon as possible.  Note that the scheme only notifies the phone user of the existence of collocation, not of the time or location of the collocation event. 
Furthermore, the public registry only holds unnamed random tokens.

While this simple scheme still has a number of weaknesses, it captures the basic approach. In the next sections we describe these weaknesses, our solutions, and a number of extensions.

\subsection{Concurrent work}
Various groups are concurrently working on various stages of solutions:
\begin{itemize}
    \itembf{China, Taiwan and South Korea} appear to be centrally aggregating cellphone tracking data to track the movements of infected and quarantined persons and notify others of potential collocation.   We are not aware of public technical details of how these systems operate.
    
     \itembf{Singapore} has very recently deployed their TraceTogether cell-phone app \cite{tracetogether} that is using Bluetooth and random identifiers akin to our proposal. Cho et al.~\cite{CIY20} provide an analysis of the privacy benefits and drawbacks of TraceTogether, noting, in particular, that TraceTogether focuses on privacy from fellow users; the system only provides limited privacy of infected users (or their collocated citizens) from the government. 

    \itembf{Israel} has also deployed their \emph{Hamagen} cellphone app~\cite{MohGovIL}, which periodically downloads to all participating phones a pseudo-anonimized list of the paths of all infected individuals. Phones then locally compare their own past locations with the infected paths. Though the app does not display the downloaded paths, the location history of infected individuals (and very possibly their identity) might be extracted from phones either by repeated querying or by jail-breaking the operating system.

    \itembf{MIT} has privacy-focused solutions~\cite{Safepaths,wsj-mit} that attempt to blur cellphone location data in order to maintain user privacy and also to perform privacy-preserving location tracing (the latter of which has similar security guarantees as the system we propose in this work).

    \itembf{Covid-Watch} group \cite{covidwatch} proposed a distributed variant of the TraceTogether approach. Based upon the description on their website as of the time of this writing, the Covid-Watch scheme may be susceptible to a linking attack (see Section \ref{sub:linking}).

\end{itemize}
In addition, Lindell and Green had an online video conversation on Brighttalk \cite{brighttalk} discussing the scientific and political challenges associated with location tracing in the context of COVID-19.
Furthermore, there exists substantial research on the related question of private proximity testing (e.g., \cite{DBLP:conf/ndss/NarayananTLHB11,DBLP:conf/percom/NielsenPS12,DBLP:conf/wisec/SaldamliCJK13,DBLP:conf/esorics/ZhengLLH12,DBLP:journals/iotj/SunZZ16}).

We note that several of these works have appeared only in the past few days, while we were in the process of writing up our approach, following our group's initial post on LinkedIn~\cite{ari-linkedin}. We apologize for any potential repetition of explanations or ideas, and we invite scrutiny and feedback from the broader community to any of the ideas in this work-in-progress document.

\section{High level structure}
\label{sec:highLevelApproach}

The scheme consists broadly of the following components:
\begin{itemize}
    \itembf{Dissemination and collection of collocation tokens.} Participating cell phones continuously broadcast special tokens over their available short-range networks; the tokens are recomputed every few minutes.   At the same time, the phones collect and store tokens broadcast by nearby phones.   Non-human ``surfaces'' or locations, such as cafe tables or meeting rooms, can participate as users by having cell-phones attached to them.
    \itembf{Reporting of infection.} When an individual is confirmed by medical personnel to be infected, the tokens disseminated during the period of potential infection are uploaded to a central registry of infected tokens.  This process will be performed in conjunction with authorized medical personnel. 
    \itembf{Notification of collocation.} Users of the scheme periodically check the tokens collected by their phones against a central registry. If a match exists then the user will know that collocation occurred, and that they should get tested.
    \end{itemize}

    The precise structure of the tokens and the details of each one of the components are of course  crucial for the  security of the scheme and will be discussed later on.  
     It is stressed, however,  that the scheme involves no registration of users or phones. Anyone can join and leave at any time.

     \section{Security requirements}
     \label{sec:requirements}
     
Before describing the basic scheme and the variants in more detail, we highlight its main functionality and security considerations.  We stress that not all variants of our scheme satisfy all the security properties listed below. Moreover, some of the privacy protections below may be obviated when the scheme is deployed in conjunction with other privacy-leaking mechanisms; for example, the privacy of a COVID-positive individual may be obviated when medical or social protocols require the individual to disclose their status.

\begin{description}
\item[Accuracy and reliability of collocation data.]
This is arguably the first and foremost security requirement: Though users are free not to participate, the scheme should not miss a collocation of an infected user with another one, as long as both use the scheme as instructed. Moreover, the scheme should guarantee a minimal false-positive rate {\em even under adversarial attacks from users or the database server.} These requirements are underscored by the fact that the notified user has no way to verify the notification, and so frequent false positives will reduce trust in the system. 
\item[Anonymity of users from other users.]
Since we insist on providing precise collocation information to individuals, we cannot always provide full anonymity. For example, if person $A$ is physically collocated only with a single person $B$ and is notified that it collocated with an infected person, then $A$ learns that $B$ is infected.  Still, we aim to guarantee that no user  can learn any information that cannot be inferred just from knowing whether they collocated with an infected user. Similarly, any coalition of users should learn only what can be deduced from the individual information that each coalition member is entitled to know.
\item[Anonymity of users from a central authority.]
We would like the central authority, including the entity managing  the registry of infected tokens, to learn nothing about the identities or locations of the participating users, whether infected or uninfected.
\item[Protection from medical personnel.]
While there is no point in providing anonymity from the medical personnel involved, we would like to guarantee that medical personnel cannot abuse the scheme to aggregate data on the location and collocation of patients.
\end{description}
\section{Technical details of the basic scheme}
We next flesh out the basic scheme  sketched in the introduction, with the aim of making the presentation accessible to readers without much technical background. Section \ref{sec:extensions} presents a number of more advanced extensions of the basic scheme.

Our system is based on three basic design parameters:
\begin{itemize}
    \itembf{Time epoch \emph{tick}} - the duration of one epoch of time in the system.  A tick should be long enough to be reliably reproduced on a variety of devices, and yet short enough to produce the desired privacy guarantees in Section~\ref{sec:analysis}.  For the purposes of exposition, the reader may think of a tick as 1 minute.
    \itembf{Retention time}  - the length of time for which information should be stored on phones and in the registry.  (For COVID-19 this period currently stands at two weeks.) 
    \itembf{Update interval $r$} - the amount of time (in ticks) between participant downloads of the registry.
\end{itemize}

\subsection{Broadcast and recording}

A participating phone continuously broadcasts its current token, which is replaced every epoch; these tokens are stored locally for prescribed retention period (\eg two weeks).  A separate process listens to broadcasted tokens from neighboring phones, and locally stores all received tokens, again for the retention period.

Tokens can be the result of a (pseudo-)random number generator or a (pseudo-)random function (such as the Advanced Encryption Standard - AES) or the result of a cryptographic hash function (such as the Secure Hash Algorithm - SHA256) applied to some more structured data (see Section \ref{sec:extensions}). 
As a back-of-the-envelope calculation, at constant activity over a two-week period (with a one-minute \emph{tick}) a phone will generate roughly 20,000 tokens. Assuming that a typical person is in close proximity to no more than 1000 people per day, at most 14,000 external tokens will stored in a two week retention period.

\subsection{Medical professional}
When a user tests positive for COVID, she will request her app to provide its list of self-generated tokens to her medical professional, who will upload the list to the registry of infected tokens.

\subsection{Registry}
The registry will include all the tokens uploaded by infected individuals. At regular intervals of $r$ ticks, the app downloads infected tokens from the registry and checks whether it has recorded any of them.  Continuing our back-of-the-envelope calculation, assuming one million infected users over a two week period, each with a record of 20,000 tokens, produces a registry of 20 billion entries.  In Sections \ref{sec:bloom} and  \ref{sec:extensions} we consider a number of ways of condensing this registry for practical implementation.
\section{Analysis}
\label{sec:analysis}

We provide an informal analysis of the properties of the basic scheme, and describe some attacks and caveats. Most of these attacks are addressed by the extensions in the extensions in Sections \ref{sec:bloom} and \ref{sec:extensions}.

\subsection{Accuracy and reliability}

First we note that the proposed scheme - as any scheme that is based on BLE transmission - will inherently have false positives that result from the fact that BLE transmission crosses walls and other commonplace separators that naturally stop infection. Such false positives can be somewhat reduced by lowering the transmission and reception energy and other means - but they can never be eliminated. Thus an alert by the scheme should always be taken as {\em potential collocation}  rather than definite one. 

In addition, we consider two attack scenarios where rogue users intentionally attempt to flood the system with false positives:
\begin{itemize}
\item
An infected user uploads a fake list of tokens to the registry. For instance, the user includes its own tokens from older time periods, or tokens that it has collected from others.  If  Bloom filters are used  (see Section \ref{sec:bloom}), the attacker can just include an all-1 filter.
\item
Rogue users collect token from ```targeted users'', and rebroadcast these tokens in multiple locations and over long periods of time.
\end{itemize}
Section \ref{sec:extensions} describes measures against both of these attacks.

\paragraph{Rogue registry.}
In the basic scheme, where all tokens reported by positive individuals are made publicly available, the only guarantee needed is authenticity and availability of the posted information. This can be guaranteed by replication, e.g.  via an existing block-chain.  In variants where the full registry is required to remain hidden from the public (see Section \ref{sec:extensions}), additional mechanisms are needed to guarantee protection against a rogue registry. 

\subsection{Privacy}

The privacy of our system is based on the various parameters of its implementation.  We divide our analysis in terms of the information leaked to four entities participating in the implementation:
\begin{itemize}
    \itembf{User} - the user of the cellphone application;
    \itembf{Registry} - the system that collects COVID status information and shares it with participating users;
    \itembf{Doctor} - the medical professional who tests the user and determines a positive COVID status, and potentially reports it to the registry; and,
    \itembf{World} - the rest of the world, which may choose to interact with the registry.
\end{itemize}

\subsubsection{Users}
\label{sec:users}

The users of such a system acquire external information through two sources:
\begin{enumerate}
    \item The tokens that are broadcast by other participants.
    \item The information obtained from the registry.
\end{enumerate}
The amount of location information gleaned about a given user is limited to the length of an epoch.  As shown below, we claim that the system provides protection against linking location information from different epochs. 

We note that a rogue user can always record additional information, such as the time, location, or neighborhood video at the time of receipt of each token.
We argue however that such attacks are ``inevitable,'' in that they are inherent to the desired functionality of the system.  The need to alert a user in a timely manner to infectious collocation restricts our ability to hide that infected individual from those in her vicinity.
The saving grace is that this attack is hard to scale -- the rogue has to be in the physical presence of the victim.  It may
also be possible for the system to identify and censure repeated attacks.  Moreover, a certain plausible deniability can be
built into the system through the use of a Bloom filter (see Section~\ref{sec:bloom}) with a non-trivial false positive rate.

\subsubsection{Registry}
\label{sub:linking}

In the basic scheme, the registry is designed to inform users about the COVID status of nearby people, all without storing sensitive data. In this way, the registry can be public and transparent without acquiring information that can be used to reidentify a participant's location trace.

In fact, we make a stronger claim: in addition to being unable to identify any user's location, the registry cannot, in general, detect whether two random tokens come from the same user or from different users. This \emph{unlinkability} promise ensures that the service cannot form a location trace of a participant's movement patterns, thereby obviating the concern that a location trace may be connected to a specific person. Note that we presume that data uploads use a network-level anonymity system like Tor (torproject.org) to avoid network-level identity leaks.

Following best practices within cryptography, we codify this strong unlinkability guarantee as follows: knowledge of the entire database cannot help an attacker learn whether a particular random token $\token$ at time $\tick$ was transmitted by either one of two users $\user_0$ or $\user_1$, even in the worst-case scenario that the server:
\begin{itemize}
    \item knows the random tokens provided by $\user_0$ and $\user_1$ at any prior or future time epoch, perhaps by observing them walking on the street; and,
    \item controls the database entries provided by any other user.
\end{itemize}

At a high level, the independence of tokens at different time epochs provides forward and backward unlinkability.  Moreover, because all collocation tests are done on the local phone, the registry is not privy to collocation results until a
participant publicizes a positive test result.

\subsubsection{Doctor}

We assume that healthcare professionals are trusted from the perspective of integrity. That is, doctors only upload tokens to the registry of a patient that is believed to have COVID.  This appears to be an unavoidable assumption of current state of affairs, where the medical world has full control over testing and result reporting to the individual.

Our protocol does, however, provide some confidentiality from doctors.
The medical professional receives a patient's generated token list, but this does not, in itself, reveal location.
Combining token lists from multiple patients does permit a doctor to learn whether her own patients have interacted, which she can (and maybe even should) learn from a proper medical history. However, a combined list from multiple patients does not permit a doctor to learn any location information about non-patients because the doctor does not known the spatial relationships between tokens.

Further protection can be afforded by having the user supply a Bloom filter of his tokens, rather than the tokens themselves (see Section~\ref{sec:bloom}.

\subsubsection{World}

Since anybody can sign up to participate in the protocol and the database is presumed to be non-sensitive, we have already captured threats against the rest of the world within the ``User'' and ``Registry'' sections above.
\section{Implementation using Bloom filters }
\label{sec:bloom}

We show how the communication and storage requirements of the system can be reduced significantly  by using Bloom filters, at the cost of adding some probability of false positives. In addition,  the use of Bloom filters adds a layer of uncertainty and ``plausible deniability'' regarding whether any particular token is included in any registry.  We consider the following two additional parameters:

\begin{itemize}
    \itembf{False positive probability \pf} - the probability that a heard token will be incorrectly considered to be from a COVID-positive person.
    \itembf{Growth bound $g$} - the maximum number of token samples to be accepted from a given COVID-positive patient.  This will affect the amount of information that the registry must share with each participant.
\end{itemize}

\subsection{Registry}
To save space, the registry will not store the infected tokens directly. Instead, it will store  an $m(t)$-bit Bloom filter of  the infected tokens, where $t$ is the number of ticks from the start of the system worldwide (or from the start of the current infection period, e.g. 14 days for COVID-19).
The length $m(t)$ of this filter is maintained by the registry, based on the number $n(t)$ of items that have
been inserted into the Bloom Filter, in order to keep the optimal false positive probability
of the filter at a designed value \pf.  For large filters~\cite{starobinski2003efficient,christensen2010new}, this means
that we should design the filter to have
\begin{equation}
    \label{eq:BloomLen}
m(t) = - \frac{n(t) \log_2 \left(\pf \right)}{ln 2} \mbox{bits},
\end{equation}
based on $k = -\log_2 \left(\pf \right)$ hashes.

Bounding the growth of the registry's Bloom filter over time could be accomplished with multiple independent filters.
The registry starts with a filter of size $s$, and adds a second filter of size $\alpha s$ as the first filter fills up (ideally well before it fills up so that testing facilities know what size filter to transmit).  When the second filter starts filling up, a third filter of size $\alpha^2 s$ is spun up, and so forth, with geometric growth.
In this way, the probability of false positive remains fixed, and the registry avoids storing the tokens it receives
from the medical professionals (which would otherwise be needed to recalculate the filter to a new size).

\subsection{Medical professional}
 The medical professional will put the tokens obtained from the patient through a Bloom filter of the appropriate size, and send the result to the registry.  In this manner, the registry never sees the tokens (and, thus, cannot correlate them to the same infected individual).  Bloom filters have the useful property that the bit-wise \emph{OR} or two filters $B_1$ and $B_2$ (containing elements $E_1$ and $E_2$ respectively) produces a resulting filter $B$ containing elements $E_1 \cup E_2$; as such, the registry can just bit-wise \emph{OR} each new medical Bloom filter report into its own master Bloom filter.

We note that if cryptographic hash functions are used in support of the Bloom filter, then the false positive probability is essentially equal to the probability of producing a fake token that matches the filter.  This means that a fake need for a COVID test can be identified with probability $1-\pf$.  To prevent replay attacks (presenting the same token to various testers), doctors would have to share Bloom filters of presented tokens with the registry as well.

\subsection{Tokens}
At each tick, the app produces and records (with a timestamp) a \emph{new} cryptographically strong random number of length $\lambda$ with probability \pnew - this is the token that is broadcast to others.
The term $\pnew$ must be tuned to the privacy analysis in Section~\ref{sec:analysis}, but also so that the expected number of tokens generated by an infected person is less than the growth bound $g$.
If the last $i$ ticks of an infected person are relevant to disease spread, then we need
\begin{equation}
\label{eq:growthBound}
\frac{i}{\pnew}
\leq g.
\end{equation}

Finally we note that multiple Bloom filters can be used to account for varying degrees of certainty that a user is infected, since COVID-19 tests have a non-zero false positive and negative error rate.

\subsection{Sample parameters}
System parameters need to be attuned to the expected size and growth of the tested population.

Assuming that the average user would be willing to use, say, $100$MB of WiFi/cellular bandwidth per day on COVID-protection, then a Bloom filter size $m(t) \leq 8x10^8$ bits
could be downloaded once per day.

To support a false positive probability of, say, $10^{-15}$, we could then provide up to $n(t) \leq 1.1x10^7$ token insertions,  based on~\eqref{eq:BloomLen}.  In a town of $10^4$ residents, with a 10\% peak infection rate, this would allow recording about $g=1000$ tokens per infected person.  Assuming that the last 20 days ($=28800$ ticks, at $1$ tick / minute) of a user's data are relevant to infection spread, this means (by~\eqref{eq:growthBound}) that
\[
\frac{1}{\pnew}\leq ~3.5\%.
\]
This means that we could have, for example, $\pnew=\frac{1}{30}$, meaning that new tokens are broadcast, in expectation, every $30$ minutes.  Of course, the Bloom filter is naturally compressible, so these are very conservative estimates.

\paragraph{Trusted server computation.}
If the server is trusted to do Bloom filter queries, then communication can be significantly curtailed, with the user only needing to send the tokens that is has seen.  Assuming, say, $100$ contacts per day, each for $5$ minutes (\ie 5 ticks), this results in communication bits $5000\lambda$ bits to the server each day, recalling that each token is of length $\lambda$ bits.  This means that even $\lambda=100$-bit tokens result in roughly $63$KB of transmission to the server.

\section{Extensions}
\label{sec:extensions}
We briefly outline several extensions to the system that improve the privacy, resiliency and features provided by the scheme -- albeit at the cost of reducing simplicity.

\subsection{Improving privacy and resilience}

\paragraph{Fine-tuning and hardening the BLE communication.}
A key to enabling this approach is naturally matching communication range to the contact distance of the disease.
There are several possible ways of doing this, including controlling the transmission power and assessing the received signal strength.

In the first case, the goal would be to make the app reduce the transmission power of the communication device (say the Bluetooth subsystem) to a point where it can be typically received in the desired range.
An adversary with a particularly sensitive or directional high-gain antenna could extend the reception radius, for the purpose of gathering IDs and maybe identifying infected parties, but even this will be limited due to the physical characteristics of the signal.

In the second case, the app could transmit at normal power, even using longer-range technologies such as WiFi, but filter out all transmissions below a prescribed received signal strength indication (RSSI) value, in effect ignoring transmissions beyond the prescribed radius.  Again, an adversary could choose to promiscuously listen to all transmissions, but the reception region would still be bounded.

\paragraph{Per-encounter tokens: Protecting from linkability and re-broadcasting.}
Instead of having cell phones directly store the tokens received from other phones, they can instead store only the combined hash of the received token and the app's generated token (at that moment of contact).   In other words, when phone $A$ receives a token $T_B$ from phone $B$, it will store $T_{AB}=H(T_A,T_B)$, where $T_A$ is the token that $A$ is currently broadcasting.($T_A$ and $T_B$ should be ordered lexicographically so as to guarantee that phones $A$ and $B$ store the same $T_{AB}$).

This method has the advantage that it reduces the ability of the registry to link collocation information coming from different individuals  (especially when the tokens uploaded by an infected individual are reordered randomly).  
Furthermore, this methods thwarts a potential ``re-broadcasting attack'',  where a malicious individual collects the tokens received by an infected (or potentially infected) individuals and replays them in many locations,  thus creating a large number of false positives.
On the other hand, this methods somewhat increases the number of false negatives relative to the basic scheme. This happens when $A$ receives $B$'s token while $B$  does not receive $A$'s token.

\paragraph{Mitigating re-identification of infected users by fellow users.}
    As discussed in Sections~\ref{sec:requirements} and~\ref{sec:users}, the functionality of the system inevitably allows a user $E$ to target  user $A$ by collocating with $A$  and then not collocating with anyone else and repeatedly checking the system for an alert.  However,  making the registry public makes such undesirable behavior simpler, undetectable, and more scalable. In particular, users that record times and locations associated with each recorded token can use this information to identify all the positive-tested users that collocated with them.  
    
    One way to mitigate such behavior is to not make the registry public, but instead to opt for having access to the registry mediated by a semi-trusted server that processes queries made by users and returns some predefined function of the collocation information. \footnote{This function can be either a binary value that represents collocation, or else a more nuanced ``risk score'' that it determined with medical advice so as to represent the level of risk of infection given the collocation history of the querying user.} Indeed, with such a registry in place, scaling the inevitable attack becomes harder and more detectable, thus allowing for potential out-of-band mitigations.  
    
    However, the introduction of a server introduces other concerns: first, the registry can obtain significantly more information this way  (for one, the server learns whether the querying party is potentially infected); second, the user needs to trust the server for correctness.  
    
   Privacy can be regained using standard cryptographic tools such as  Private Information Retrieval (PIR) and  Private Set Intersection (PSI) to allow the user to obtain the needed information without having the server learn anything at all.  Authenticity can also be guaranteed using standard tools such as zero-knowledge proofs, or more specifically zero-knowledge sets that allow for efficient, privacy preserving proof of membership and non-membership in a set. 
   
\paragraph{Verifying physical proximity.} A multi-message handshake between two nearby users can support proximity, helping to thwarting some remote relay attacks at the expense of a more complex communication protocol.
     
\paragraph{Verifying token validity.}
The ability of rogue infected users to upload to the registry fake (say, replayed) tokens can be thwarted by having the token be a one-way hash of another random identifer  that is linked to the patient or the phone. The medical personnel can then verify the validity of the tokens uploaded to the registry.
    
    \paragraph{Preventing rogue re-broadcasting of tokens (take II).}
    Another way to prevent rogue re-broadcasting of tokens is to have the tokens include a hash of values that can be verified at the time of collocation, such as  curent time  or GPS location. This allows the collocated parties to verify the freshness of the tokens before registering them locally.  Clearly, for privacy reasons, the time and location information will not be kept  - it is no longer needed once the verification is done.
    
    \paragraph{Enabling anonymity and verifiability when querying the registry.}
    A private publish-subscribe system (e.g., \cite{DBLP:conf/nss/CrescenzoBCSSTW13,DBLP:conf/isorc/KhouryLPTL14}) may allow users to be notified when they come into contact with a person who later tests positive for COVID-19, without the need to download the entire registry. Informally, such a system would permit users to register subscriptions for all random tokens that they received, so that they are notified if an infected patient later publishes the same token.

\subsection{Additional features}

\paragraph{Detecting staggered collocation.} 
The system can be augmented by attaching devices to locations (such as public surfaces) that register as ``contagious'' once they hear a contagious token.  The devices can participate in the system in the same way as users, except that they will report as infected only tokens that were broadcast until the point in time where the surfaces were cleaned or otherwise stopped being contagious.

\paragraph{Adding update capability.}
An update mechanism can be added to the system to enable subsequent features or privacy protections. Normally, privacy-preserving protocols lack the ability to be updated since the legacy data is protected in a manner that only permits the acceptable computations. In this scenario, all data is only valid for 14 days or so, after which it is acceptable to ``hard fork'' to an updated privacy mechanism. Any change should be made transparently, with informed consent from all existing users, and this can be enforced through the app update mechanism of the phone.
\section{Conclusions}
Several elements are key to the success of the proposed system.
\paragraph{Preventing abuse.}
It should not be possible to inundate the system with fake IDs or prevent users from accessing the registry.

    \paragraph{Maintaining authenticity.}
    It is essential that a broad cross-section of society can reliably access and install the app, without fear of installing counterfeit copies, or accessing non-authoritative data.
    
    \paragraph{Adoption.}
    Perhaps by far the greatest hurdle for this app is adoption - very quickly getting a large body of people to use the application, including medical professionals who are administering tests.  This would require support from a wide cross-section of national resources, which we hope that this publication will facilitate.
    
    One possible inducement would be to make observed IDs serve as tickets to faster COVID-testing.  A participant who could show her doctor a valid ID from an infected individual as evidence of her need for a test.  In a status quo of insufficient test availability, this could be an attractive option.
    A stronger inducement could be to simply pay participants for producing correct certificates for the need to be tested.  Opportunistic risk of contracting the virus will be mitigated by (i) the fear of actually getting sick, and (ii) the subsequent need for a 14-day quarantine.

\section*{Acknowledgments}

This work is an offshoot of a discussion, initiated by Andy Sellars,  on the privacy impacts of COVID-containment efforts. Thanks Andy!  We also gratefully acknowledge the helpful conversations and comments from Gerald Denis, Anand Devaiah, Amir Herzberg, David Starobinski, and Charles Wright. Finally, we are grateful for the productive discussions with  Ramesh Raskar, Ron Rivest, and the rest of the MIT Safepaths team \cite{Safepaths}.

\bibliographystyle{plain}
\bibliography{ms}

\end{document}